\documentclass[aps,amsmath,amssymb,twocolumn]{revtex4}
\usepackage[]{graphicx}
\usepackage{color}
\newcommand{\be}{\begin{equation}}
\newcommand{\ee}{\end{equation}}
\newcommand{\ben}{\begin{eqnarray}}
\newcommand{\een}{\end{eqnarray}}
\newcommand{\bes}{\begin{subequations}}
\newcommand{\ees}{\end{subequations}}

\newcommand{\nn}{\nonumber\\}
\newcommand{\bfi}{\begin{figure}}
\newcommand{\efi}{\end{figure}}
\newcommand{\bc}{\begin{center}}
\newcommand{\ec}{\end{center}}
\newcommand{\sech}{\mbox{sech}}

\begin{document}
\title{Models of modified $F(R,T)$ and cuscuton braneworld}
\author{D. Bazeia$^{1}$ and A. S. Lob\~ao Jr.$^{2}$}
\affiliation{$^1$Departamento de F\'\i sica, Universidade Federal da Para\'\i ba, 58051-970 Jo\~ao Pessoa, PB, Brazil\\
$^2$Escola T\'ecnica de Sa\'ude de Cajazeiras, Universidade Federal de Campina Grande, 58900-000 Cajazeiras, PB, Brazil}
\begin{abstract}
This work deals with braneworld models in the presence of scalar fields in a five-dimensional warped geometry with a single extra dimension of infinite extent. We consider generalized models in the presence of the Ricci scalar, the trace of the stress-energy tensor and the cuscuton contribution. The models describe novel braneworld scenarios and the investigations consider distinct possibilities, from which we show how the brane may change to become thinner, although keeping gravitational stability and the gravity zero mode under strict control. Moreover, we did not identify any split behavior in the warp factor in the presence of the cuscuton and the trace of the stress-energy tensor.
\end{abstract}
\maketitle

{ {\bf 1. Introduction.} --} About two decades ago, it was theorized that the four-dimensional universe could be embedded in a higher dimensional spacetime denoted by bulk \cite{RS,GW}. The proposed theory became known as braneworld models, and important contributions have been presented to intriguing problems in physics such as the origin of dark energy \cite{Csaki:1999mp} and cosmological inflation \cite{Schwindt:2005ns}, as results of these studies. In this scenario, theories that differ from standard General Relativity, so-called modified gravity theories, have attracted much attention recently since they present new alternatives to describe specific problems. In the braneworld scenario, real scalar fields have been used for descriptive topological structures in an $AdS_5$ geometry with a single extra dimension of infinite extent, inducing the location of brane in a domain wall \cite{DeWolfe:1999cp,Gremm:1999pj,Csaki:2000fc}. Thus, models that engender topological structures have been widely studied by several authors with distinct motivations; see, for example, \cite{ref1,ref2,ref4,ref5,Bazeia:2013uva,CE,liu,olmo215011,Souza,CE1} and references therein.

Among the theories of modified gravity \cite{R1,R2}, some of them proposed as $F(R,T)$, where $R$ represents the Ricci scalar and $T$ the trace of the stress-energy tensor \cite{Harko:2011kv},  call our attention as they allow topological structures in braneworld scenarios  \cite{ , Bazeia:2015owa,Gu:2016nyo, Bazeia:2013oha, Afonso:2007zz}. Models with modification in the interaction and dynamics of scalar fields have also presented new characteristics such as split behavior \cite{Bazeia:2015owa} and compactification of the brane \cite{Bazeia:2015eta}. Recently, brane models with modified dynamics with the cuscuton term were investigated in the context of standard gravitation leading to new split behavior in warp factor \cite{Andrade:2018afh,Bazeia:2021jok}. In Ref. \cite{Bazeia:2021jok}, in particular, the cuscuton term was shown to change significantly the warp factor of the brane, opening up the geometric behavior of the brane into the extra dimension. A similar modification was found in \cite{Lobo}, but in a different scenario where $T$ is absent and cuscuton is part of the souce Langrange density. This new possibility motivated the present study, which differs from previous research since now both the cuscuton and the trace $T$ are present in the braneworld models described by a single real scalar field.

This paper aim to investigate braneworld scenarios in the presence of kinklike solutions in  generalized models of thick brane with $F(R,T)$ in the form $F(R)+G(T)$, with the inclusion of the cuscuton term to the standard kinematics of the scalar field. The work is outlined as follows: Sec. 2 introduces the general formalism that describes the proposed model in the presence of a single extra dimension of infinite extent, and investigates the linear stability. Sec. 3 describes two analytical solutions for the case $F(R)=R$ and $G(T)=\beta T$, using the first order formalism. Still in Sec. 3, a more general model described by $G(T)=\beta T$ and $F(R)=R-f(\beta) R^2$ is studied. Lastly, Sec. 4 presents the conclusion and some comments on future possible investigations derived from the present study. \\

{\bf 2. General Model.} -- Let us consider an action for 5$-$dimensional spacetime which describes a generalized gravity written as,
\begin{equation}\label{eq1}
    S=\!\int d^{5}x \sqrt{|g|\,}\Big[-\frac14 F(R)-G(T)+{\cal L}\,\Big],
\end{equation}
where $F(R)$ and $G(T)$ in general represent functions of the Ricci scalar $R$ and the trace $T$ of the stress-energy tensor of the source scalar field. The Lagrange density engenders the cuscuton term, whose strength is controlled by $\alpha$ factor in the form,
\begin{equation}\label{eq2}
    {\cal L}=\frac12\nabla_a\phi\nabla^a\phi+\alpha \sqrt{|\nabla_a\phi\nabla^a\phi|}-V(\phi)\,.
\end{equation}
We are using natural units and $4\pi G_5 = 1$, $g = det(g_{ab})$, $a,b = 0,1,...,4$ and the signature of the metric is $(+ - - - -)$. Furthermore, we investigate the flat brane case with the line element given by $ds^2 =e^{2A}\eta_{\mu\nu}dx^\mu dx^\nu-dy^2$ where $e^{2A}$ is the warp factor, $\eta_{\mu\nu}$ is the 4$-$dimensional Minkowski metric, and the Greek indices for the embedded $(3+1)-$dimensional space, $\mu, \nu =0,1,2,3$ with the extra dimension defined as $y=x^4$.

We can use the source Lagrange density \eqref{eq2} to write the stress-energy tensor
\begin{equation}\label{eq3}
    T_{ab}=\Big(1+\alpha\frac{\sqrt{|\nabla_c\phi\nabla^c\phi|}}{\nabla_c\phi\nabla^c\phi}\Big)\nabla_a\phi\nabla_b\phi-g_{ab}{\cal L}\,,
\end{equation}
which allows us to calculate the trace $T=g^{ab}T_{ab}$ as,
\begin{equation}\label{eq4}
    T=-\frac32\nabla_a\phi\nabla^a\phi-4\alpha \sqrt{|\nabla_a\phi\nabla^a\phi|}+5V(\phi)\,.
\end{equation}

The equation of motion for the scalar field $\phi$ can be found by varying the action associated to the Lagrange density \eqref{eq2} with respect to the scalar field, and can be written as,
\ben
\!\!\!&&\!\!\!\nabla_a\Big[\Big(1+\alpha\frac{\sqrt{|\nabla_c\phi\nabla^c\phi|}}{\nabla_c\phi\nabla^c\phi}\Big)\nabla^a\phi\Big]+{V}_\phi+5G_T{V}_\phi+\nn
\!\!\!&&\!\!\!+\nabla_a\Big[G_T\Big(3+4\alpha\frac{\sqrt{|\nabla_c\phi\nabla^c\phi|}}{\nabla_c\phi\nabla^c\phi}\Big)\nabla^a\phi\Big]=0\,,\label{eq5}
\een
where $V_\phi=dV/d\phi$ and $G_T=dG/dT$. In addition, the Einstein equation arising from the variation of action with respect to metrics $g_{ab}$,
\ben
\!\!\!\!\!&&\!\!\!\!\!F_R R_{ab}-\frac12 g_{ab}F+\Big(g_{ab}\Box-\nabla_a\nabla_b\Big)F_R=2T_{ab}+\nn
\!\!\!\!\!&&\!\!\!\!\!+\,2g_{ab}G+6G_T\Big(1\!+\!\frac43 \alpha\frac{\sqrt{|\nabla_c\phi\nabla^c\phi|}}{\nabla_c\phi\nabla^c\phi}\Big)\!\nabla_a\phi\nabla_b\phi\,,\label{eq6}
\een
where $F_R=dF/dR$. As usual, we assume that the scalar field $\phi$ and the warp function $A$ depend only on the extra dimension, i.e, $\phi=\phi(y)$ and $A=A(y)$. This condition allows us to write the equation of motion \eqref{eq5} as,
\ben\label{eq7}
\!\!\!\!\!&&\!\!\!\!\!(1+3G_T)\phi^{\prime\prime}+4A^{\prime}\Big[(1+3G_T)\phi^{\prime}-\alpha(1+4G_T)\Big]\!+\nn
\!\!\!\!\!&&\!\!\!\!\!-\,G_{TT}(4\alpha-3\phi^{\prime})\Big[5\phi^{\prime}V_\phi+(3\phi^{\prime}-4\alpha)\phi^{\prime\prime}\Big]\!+\nn
\!\!\!\!\!&&\!\!\!\!\!-(1+5G_T)V_\phi=0\,,
\een
where $\phi^{\prime}=d\phi/dy$, $\phi^{\prime\prime}=d^2\phi/dy^2$ and $A^{\prime}=dA/dy$. Also, the non-vanishing components of Einstein's equations for static solutions are
\bes\label{eq8}
\ben
\!\!\!&&\!\!\!A^{\prime\prime}F_R+\frac13A^{\prime}F_R^{\prime}+\frac13F_R^{\prime\prime}=-\frac23(1+3G_T)\phi^{\prime 2}+\nn
\!\!\!&&\!\!\!+\,\frac23\alpha(1+4G_T) \phi^{\prime}\,,\label{eq8a}\\
\!\!\!&&\!\!\!2(A^{\prime 2}\!+\!A^{\prime\prime})F_R\!-\!\frac14F\!-\!2A^{\prime}F_R^{\prime}\!=\!-\frac12(1\!+\!6G_T)\phi^{\prime 2}\!+\nn
\!\!\!&&\!\!\!+\,V(\phi)+G+4\alpha G_T \phi^{\prime}\,.\label{eq8b}
\een
\ees
We can also get the Ricci and the Kretschmann $(K)$ scalars and the trace of the stress-energy tensor as
\ben\nonumber
\!\!\!&&\!\!\!R=20A^{\prime 2}+8A^{\prime\prime}\,,\nonumber\\
\!\!\!&&\!\!\!K=40A^{\prime 4}+16A^{\prime\prime 2}+32A^{\prime 2}A^{\prime\prime}\,,\nonumber\\
\!\!\!&&\!\!\!T=\frac32\phi^{\prime 2}-4\alpha\phi^{\prime}+5V(\phi)\,.\nonumber
\een
These quantities are important for delimiting the values of parameters that define the model and cannot present singularities along the extra dimension. An important characteristic of the brane is its energy, which is given by
\begin{equation}\label{eq9}
    E=\int dy\,e^{2A}{\left(\frac12 \phi^{\prime2}-\alpha \phi^\prime+V(\phi)\right)}\,.
\end{equation}
In standard theory, the warp factor vanishes exponentially, asymptotically, and the total energy in general integrates to zero, leading to stability of the models. Here, the inclusion of the term $G(T)$ leads us to a new behavior, so we must check if the modification proposed above contribute to destabilize the geometric degrees of freedom of the braneworld model. Bellow we will investigate this issue studying linear stability of the gravity sector.

{\it 2.1. Linear Stability.}
In braneworld model the investigation of linear stability is done assuming that the metric and the scalar field undergo small fluctuations in the form $
\eta_{\mu\nu}\to \eta_{\mu\nu}+h_{\mu\nu}(r,y)$ and $\phi\to\phi(y)+\xi(r,y)$,
where $r$ represents the four-dimensional position vector. The tensor $h_{\mu\nu}$ satisfies the transverse and traceless (TT) conditions $\partial^{\mu} h_{\mu\nu}=0$ and $h=\eta^{\mu\nu}h_{\mu\nu}=0$. These conditions simplify the investigation of stability so that we can write the equation of motion to the fluctuation $h_{\mu\nu}$ as
\ben\label{eq10}
\Big[-\partial_y^2-4A^\prime\partial_y+e^{-2A}\Box^{(4)}-\frac{F_R^{\prime}}{F_R}\partial_y\Big]h_{\mu\nu}=0,
\een
where $\partial_y=d/dy$ and $\partial_y^2=d^2/dy^2$.

Considering a new z-coordinate, such that $dz=e^{-A(y)}dy$, it is possible to show that the metric become conformally flat, and the previous equation could be written as a Schroedinger-like equation in the form,
\ben\label{eq11}
\Big(-\frac{d^2}{dz^2}+U(z)\Big)\bar{h}_{\mu\nu}=p^2\bar{h}_{\mu\nu}\,,
\een
where $h_{\mu\nu}(r,y)=e^{-ipr}e^{-3A(z)/2}F_R^{-1/2}\,\bar{h}_{\mu\nu}$ and
\ben\label{eq12}
U(z)\!\!\!&=&\!\!\!\frac{9}{4} A_z^2 + \frac32 A_{zz}+\frac32A_z \frac{d(\ln F_R)}{dz}-\nn
\!\!\!&&\!\!\!-\frac14 \Big[\frac{d(\ln F_R)}{dz}\Big]^2+\frac1{2F_R}\frac{d^2 F_R}{dz^2}\,.
\een
Here one considers that $A_z=dA/dz$ and $A_{zz}=d^2A/dz^2$. With a little algebra one can use $U(z)$ to write
\ben\nonumber
S^{\cal y} S=-\frac{d^2}{dz^2}+U(z)\,,
\een
where
\ben\nonumber
S^{\cal y}=-\frac{d}{dz}+\frac32A_z+\frac12\frac{d}{dz}\Big(\ln F_R\Big)\,.
\een
Since $S^{\cal y} S$ is non-negative, the gravity sector is then linearly stable. This result was shown before in Ref. \cite{Bazeia:2015owa} and ensures that despite the modification introduced in the gravity sector of braneworld, the model is linearly stable. As the warp function depends on both the cuscuton and the trace terms, it will modify the profile of the potential \eqref{eq12} as we show below in some specific situations that allows the presence of kinklike configurations.

We can also calculate the massless graviton state, represented by the zero mode in the form,
\ben\nonumber
H_0(z)=N_0 e^{3A/2}F_R^{1/2}\,,
\een
where $N_0$ is a normalization factor that can be obtained by integration of the zero mode
\ben\nonumber
N_0^2\int e^{2A(y)}F_R\,dy=1\,.
\een

{\bf 3. Specific Models.} -- Let us now investigate some specific models, considering two distinct possibilities to compose $F(R)$ and $G(T)$. \\

{\bf 3.1. First Model.} --
In order to obtain analytical solutions, we will first consider the case where $F(R)=R$ and $G(T)=\beta\,T$, where $\beta$ is a parameter that controls the presence of $T$. With these choices, the equation of motion \eqref{eq7} becomes,
\ben\label{eq13}
\phi^{\prime\prime}+4A^{\prime}\left(\phi^\prime-\alpha\kappa\right)-\lambda V_\phi=0\,,
\een
where $\kappa=(1+4\beta)/(1+3\beta)$ and $\lambda=(1+5\beta)/(1+3\beta)$. In this work, we consider $\beta>-1/5$ to avoid problem with the braneworld model. The Einstein equations \eqref{eq8} become,
\bes\label{eq14}
\ben
A''\!\!\!&=&\!\!\!-\frac23(1+3\beta)\phi^{\prime 2}+\frac23\alpha(1+4\beta) \phi^{\prime}\,,\label{eq14.1}\\
A'^2\!\!\!&=&\!\!\!\frac16(1+3\beta)\phi^{\prime 2}-\frac13(1+5\beta)V\,.\label{eq14.2}
\een
\ees
One can now implement the first order formalism. After some mathematical manipulations, we have found the potential
\begin{equation}\label{eq15}
    V(\phi)=\frac{1}{2\lambda}\Big[\frac{W_\phi}{1+3\beta}+\alpha\kappa\Big]^2-\frac43\,\frac{W^2}{1+5\beta}\,,
\end{equation}
where $W\!=\!W(\phi)$ is an auxiliary function which depends only on the scalar field $\phi$. In this case, we obtain the following first order differential equations
\begin{equation}\label{eq16}
    \phi^{\prime}=\frac{W_\phi}{1+3\beta}+\alpha\kappa\,;\qquad A^\prime=-\frac23 W\,.
\end{equation}
We have checked that solutions of the above first order equations solve the equations of motion of the braneworld model. In particular, we observe that both the cuscuton (by way of $\alpha$ parameter) and the trace $T$ (by way of $\beta$ parameter) contribute to specify the solution of the above first order equation for the scalar field, and this also change the warp factor through the above first order equation for the warp function $A$. 

For this model the energy density can be written as
\ben
\rho(y)\!\!&=&\!\!\frac{\,\,\left(e^{2A}W\right)^\prime}{1+5\beta}+\frac{\beta\, e^{2A}W_\phi^2}{\lambda(1+3\beta)^3}+\nn
\!\!&&\!\!+\frac{\alpha\beta\, e^{2A}}{(1+3\beta)^2}\Big(W_\phi\!+\!\frac{\alpha\beta\kappa}{\lambda}\Big)\,.
\een
We notice that the energy density can no longer be written as a total derivative, however we can still get a total zero energy depending on the specific choice of parameters. Furthermore, if $\beta=0$, we return to the results obtained in \cite{Bazeia:2021jok}, as expected. Thus, with the above general procedure, let us now consider two specific cases for the $W$ function that allows interesting solutions.

{\it 3.1.1. Scenario A. } In this scenario we assume that
\ben\label{eq17}
W(\phi)=(1+3\beta)\Big(\phi-\frac13\phi^3\Big)-\alpha(1+4\beta)\phi\,.
\een
This is a simple generalization of the well-known potential $\phi^4$ that allows spontaneous symmetry breaking in classical theories in flat spacetime \cite{Babichev:2006cy,Bazeia:2007df,Bazeia:2008tj}. For this choice, the field solution have the standard shape $\phi(y)=\tanh(y)$ that provides asymptotic values of $y\to\pm\infty$ as $\phi_{\pm}\to\pm1$. We can also obtain the scalar potential using equations \eqref{eq17} and \eqref{eq15} in the form
\ben
V(\phi)\!&=&\!\frac1{54\lambda}\left(27-18c_1\phi^2+c_2\phi^2-8c_3\phi^6\right)\,,
\een
where the constants are $c_1=7+12\beta-4\alpha(1+4\beta)(2-\alpha\kappa)$, $c_2=75+144\beta-48\alpha(1+4\beta)$ and $c_3=1+3\beta$.

It is known that the asymptotic behavior of the scalar potential defines the five-dimensional cosmological constant. For the model proposed here it is obtained that
\ben\label{eq19}
\Lambda_5\equiv V(\phi_{\pm})=-\frac4{27\lambda^2}(1+5\beta)(2-3\alpha\kappa)^2\,.
\een
Note that if $\alpha\kappa=2/3$ we have $\Lambda_5=0$, showing that the bulk is asymptotically Minkowski. On the other hand, since $\beta>-1/5$ one has $\Lambda_5<0$, which makes the bulk asymptotically $AdS_5$. Using the first order equations with the standard solution for the scalar field one can find the warp function
\ben\label{eq20}
A(y)\!\!\!&=&\!\!\!\frac29(1+3\beta)(2-3\alpha\kappa)\ln[\sech(y)]+\nn
\!\!\!&&\!\!\!-\frac19(1+3\beta)\tanh^2(y) \,.
\een

\begin{figure}[!htb]
    \begin{center}
        \includegraphics[scale=0.6]{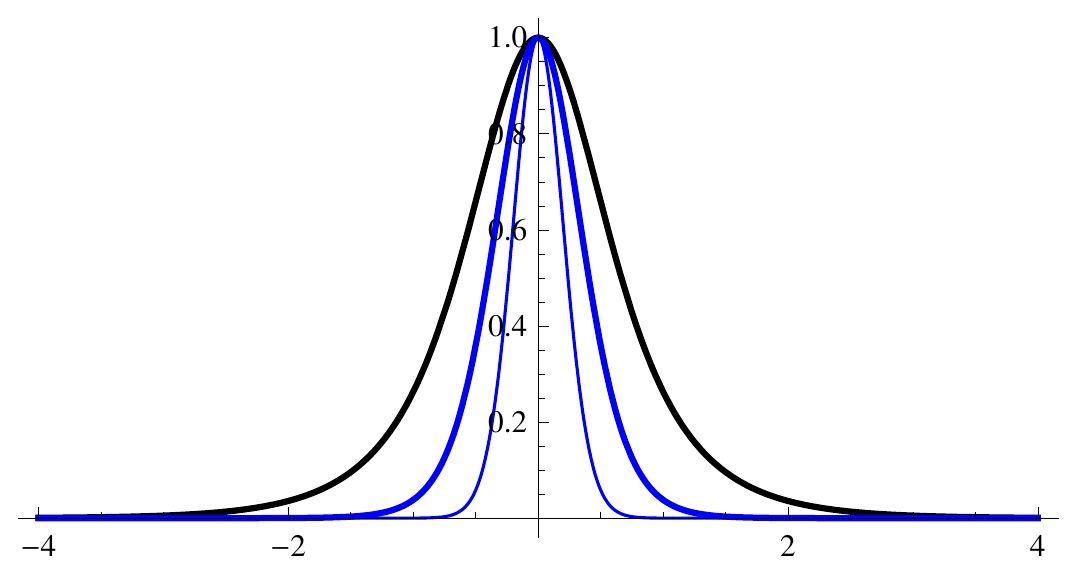}
    \end{center}
    \vspace{-0.7cm}
    \caption{\small{Warp factor for $\alpha\!=\!1/4$ and $\beta\!=\!1$ (black line), $\beta\!=\!3$ and $9$ (blue line). The line thickness decrease as $\beta$ increases.\label{fig1}}}
\end{figure}

To ensure localization of the brane we must impose that warp factor $e^{2A(y)}$ vanishes asymptotically. This condition impose new restrictions on the parameters in the form $3\alpha\kappa<2$. Fig. \ref{fig1} shows the warp factor for $\alpha\!=\!1/4$ and $\beta\!=\!1$ (black line), and also $\beta\!=\!3$ and $9$, with the thickness of the lines decreasing as $\beta$ increases. Note that when $\beta$ increases the warp factor becomes thinner and thinner, getting more and more concentrated near the origin. We can calculate the width of the warp factor at half-height to get the values $0.666$, $0.405$ e $0.238$, for $\alpha=1/4$ and $\beta=1$, $3$ and $9$, respectively. The calculation here is numerical, and we have also checked that for $\alpha=0$ and $\beta=0$, the width gets back to the standard braneworld scenario with the $\phi^4$ potential. Also, for $\alpha=1/4$ and for larger and larger values of $\beta$, the width diminishes towards zero.   

For this model one can see that there is no singularity in both the Ricci and Kretschmann scalars. In the limits $y\to\pm\infty$, the Ricci and Kretschmann scalars become
\ben
\lim_{y\to\pm\infty}R\!\!&=&\!\!\frac{80}{81}(1+3\beta)^2(2-3\alpha\kappa)^2\,,\nn
\lim_{y\to\pm\infty}K\!\!&=&\!\!\frac{640}{6561}(1+3\beta)^4(2-3\alpha\kappa)^4\,.\nonumber
\een
On the other hand, near the origin we get that
\ben
\lim_{y\to0}R\!\!&=&\!\!-\frac{16}{3}(1+3\beta)(1-\alpha\kappa)\,,\nn
\lim_{y\to0}K\!\!&=&\!\!\frac{64}{9}(1+3\beta)^2(1-\alpha\kappa)^2\,.\nonumber
\een
In addition, the trace $T$ behaves as
\ben
\lim_{y\to\pm\infty}T\!\!&=&\!\!-\frac{20}{27\lambda^2}(1+5\beta)(2-3\alpha\kappa)^2\,,\nn
\lim_{y\to0}T\!\!&=&\!\!\frac{4+15\beta}{1+5\beta}-4\alpha\,.\nonumber
\een
These quantities behave nicely for the values of parameters which we use in this work.
Differently from the proposal presented in \cite{Bazeia:2020qxr} with the cuscuton term, the warp factor do not split in the present model. However, we do not know if this change is due to the inclusion of the trace term or if it is due to the absence of the second field which appeared in the model studied in \cite{Bazeia:2020qxr}.

\begin{figure}[!htb]
    \begin{center}
        \includegraphics[scale=0.6]{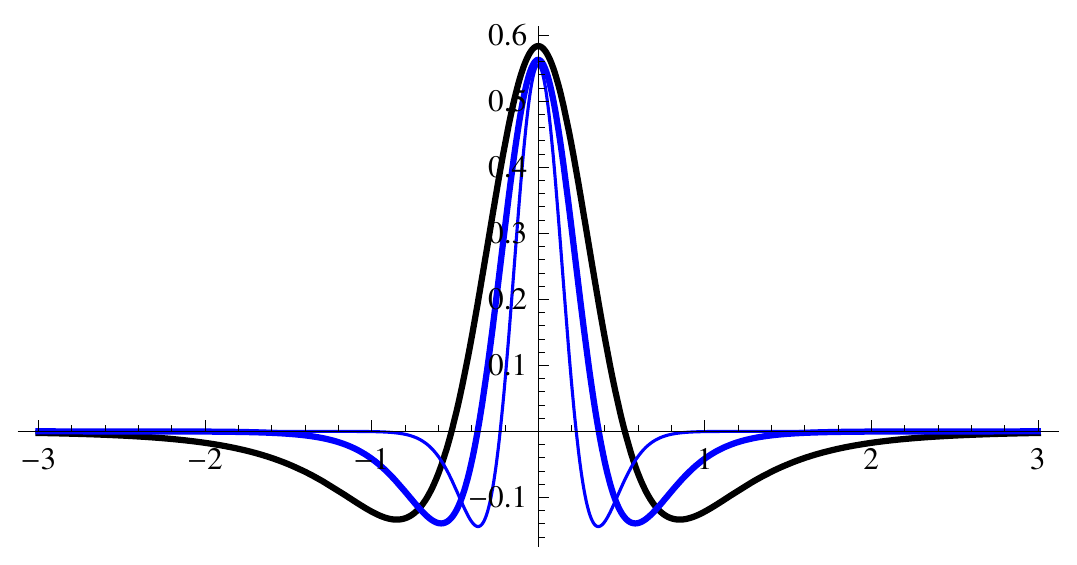}
    \end{center}
    \vspace{-0.7cm}
    \caption{\small{Energy density for for $\alpha\!=\!1/4$ and $\beta\!=\!1$ (black line), $\beta\!=\!3$ and $9$ (blue line). The line thickness decrease as $\beta$ increases.\label{fig2}}}
\end{figure}

\begin{figure}[!htb]
    \begin{center}
        \includegraphics[scale=0.6]{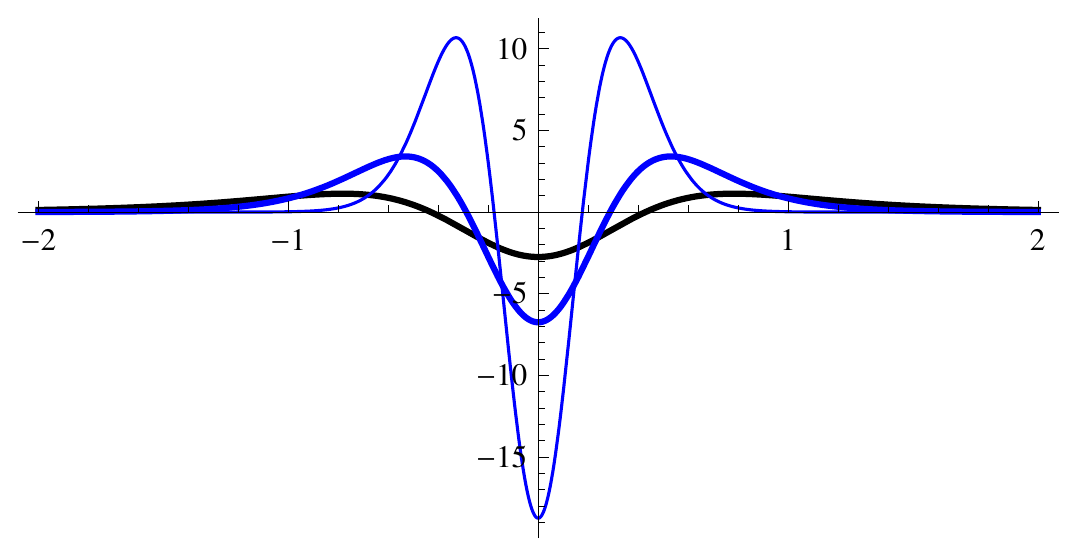}
    \end{center}
    \vspace{-0.7cm}
    \caption{\small{Potential of gravitational mode for $\alpha\!=\!1/4$ and $\beta\!=\!1$ (black line),  $\beta\!=\!3$ and $9$ (blue line). The line thickness decrease as $\beta$ increases.\label{fig3}}}
\end{figure}

\begin{figure}[!htb]
    \begin{center}
        \includegraphics[scale=0.6]{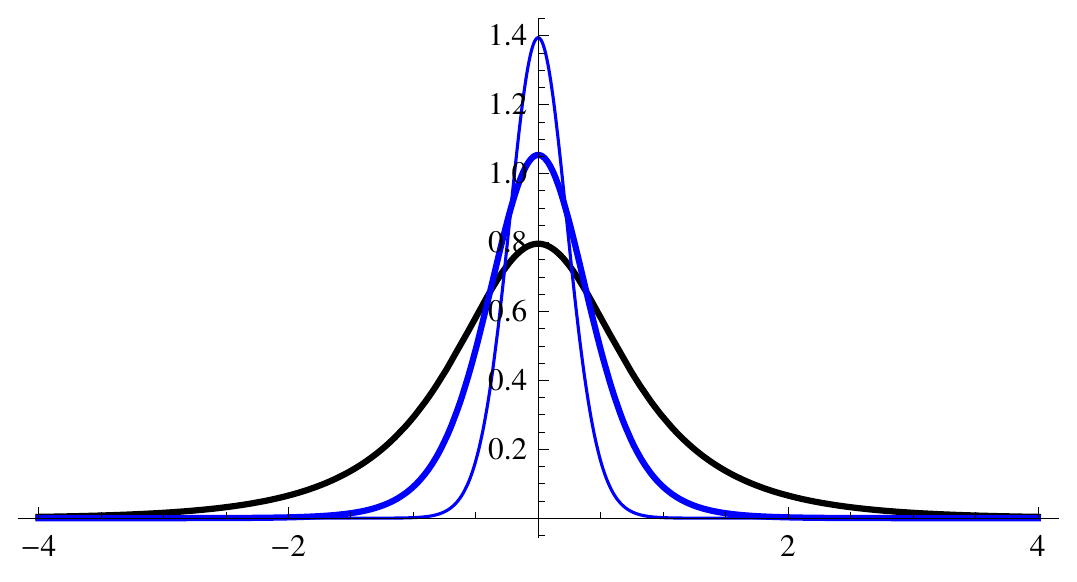}
    \end{center}
    \vspace{-0.7cm}
    \caption{\small{Zero mode for $\alpha\!=\!1/4$ and $\beta\!=\!1$ (black line),  $\beta\!=\!3$ and $9$ (blue line). The line thickness decrease as $\beta$ increases.\label{fig4}}}
\end{figure}

The energy density is represented in Fig. \ref{fig2} for the same values of parameters $\alpha$ and $\beta$ used to calculate the warp factor. Figs. \ref{fig3} and \ref{fig4} show the potential of gravitational and zero mode obtained numerically for the same parameters described above. In all cases, the $\beta$ parameter may vary to shrink the configuration toward the origin of the extra dimension.


{\it 3.1.2. Scenario B. }
Let us now consider other specific scenario, which also supports kinklike structures defined by $W$ in the form
\ben\label{eq22}
W(\phi)=(1+3\beta)\sin(\phi)\,,
\een
which is a sine-Gordon type model already studied in the literature. In the preset case, however, the first order equations allow solutions in the form
\ben\label{sol2}
\phi(y)\!=\!2\arctan\!\left[\frac{1\!+\!\alpha\kappa}{\sqrt{1\! -\! (\alpha\kappa)^2}}\tanh\Big( \frac12\!\sqrt{1\!-\!(\alpha\kappa)^2} \,y\Big)\!\right]\!,
\een
where $\alpha\kappa$ is defined in the interval $-1<\!\alpha\kappa\!<1$ to ensure that the solution has the standard kinklike profile, that asymptotically assumes the constant values,
\ben\nonumber
\lim_{y\to\pm\infty}\phi(y)=\pm \,2\arctan\!\left(\frac{1\!+\!\alpha\kappa}{\sqrt{1\! -\! (\alpha\kappa)^2}}\right)\,.
\een

We can use the above $W$ to write the Potential \eqref{eq15} as
\ben\label{eq24}
V(\phi)=\frac1{6\lambda}\left(s_1+6\alpha\kappa\cos(\phi)+s_2\cos^2(\phi)\right)\,,
\een
where the constants are $s_1=3(\alpha\kappa)^2-8(1+3\beta)$ and $s_2=11+24\beta$. Therefore, the five-dimensional cosmological constant is
\ben\label{eq25}
\Lambda_5=-\frac4{3\lambda^2}(1+5\beta)[1-(\alpha\kappa)^2]\,.
\een
For this model the warp function is
\ben\label{eq26}
A(y)\!=\!-\frac23(1+3\beta)\ln\!\left[\frac{\alpha\kappa-\cosh(\sqrt{1\!-\!(\alpha\kappa)^2}\,y)}{\alpha\kappa-1}\right].
\een
We checked that the warp factor $e^{2A}$ goes asymptotically to zero for the range of values which we are using in this work. 

\begin{figure}[!htb]
    \begin{center}
        \includegraphics[scale=0.6]{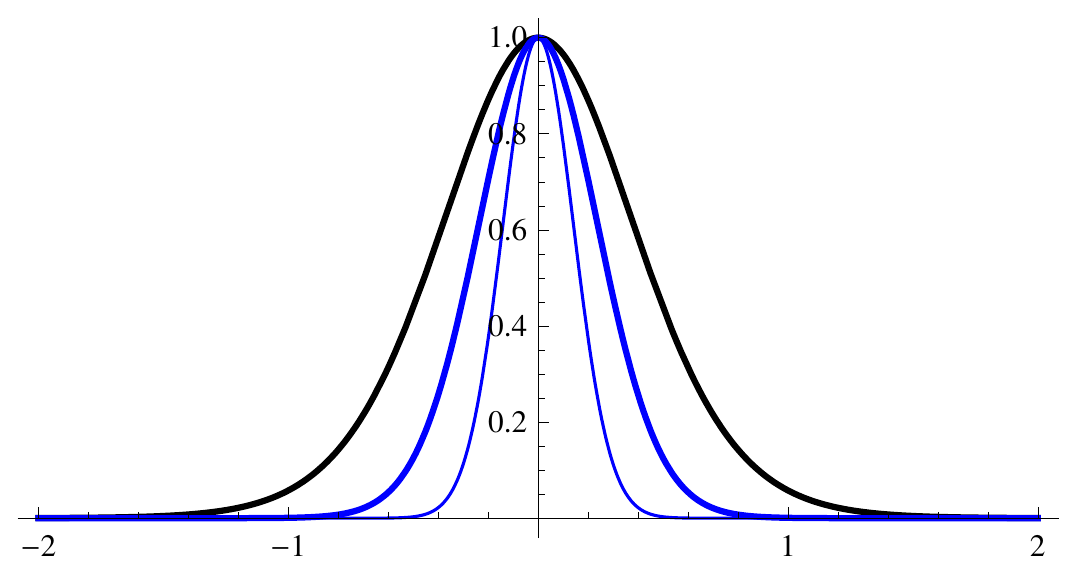}
    \end{center}
    \vspace{-0.7cm}
    \caption{\small{Warp factor for $\alpha\!=\!1/4$ and $\beta\!=\!1$ (black line), $\beta\!=\!3$ and $9$ (blue line). The line thickness decrease as $\beta$ increases.\label{fig6}}}
\end{figure}

Fig. \ref{fig6} shows the warp factor for $\alpha\!=\!1/4$, $\beta\!=\!1$ (black line), and also $\beta\!=\!3$ and $9$, with the thickness of the lines decreasing as $\beta$ increases. As in the previous result the $\beta$ parameter acts to modify the width of the warp factor. We can also calculate the width at half-height of the warp factor to get the analytical result
\begin{equation}
L= \frac{2}{\sqrt{1\!-\!(\alpha\kappa)^2}}\,  {\rm arctanh}\Big[\!\frac{1\!-\!\alpha\kappa}{\sqrt{1\!-\!(\alpha\kappa)^2}}\tan\left(\Theta\right)\!\Big]\,,
\end{equation}
where
\begin{equation} \nonumber
    \Theta=\frac12\arccos\bigl[(1+\alpha\kappa)8^{-1/(4+12\beta)}-\alpha\kappa\bigr]\,.
\end{equation}
Using the values $\alpha=1/4$, and $\beta=1$, $3$ and $9$, considered in Fig. \ref{fig6}, we get $L=0.456$, $0.283$ and $0.168$ respectively. If $\beta$ increases to larger and larger values, $L$ diminish towards zero. For $\alpha=0$ and $\beta=0$ we get back to the standard braneworld scenario with the sine-Gordon potential; we then see that these parameters may be used to control the thickness of the brane in this new scenario, in the presence of $\alpha$ and $\beta$, which control the cuscuton and the trace of the stress-energy tensor.

\begin{figure}[!htb]
    \begin{center}
        \includegraphics[scale=0.6]{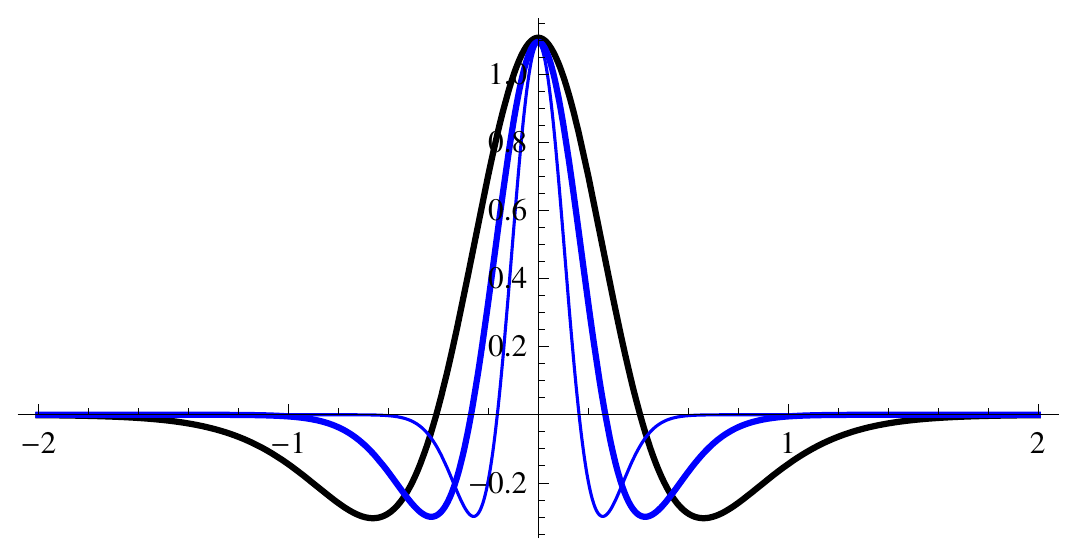}
    \end{center}
    \vspace{-0.7cm}
    \caption{\small{Energy density for $\alpha\!=\!1/4$ and $\beta\!=\!1$ (black line), $\beta\!=\!3$ and $9$ (blue line). The line thickness decrease as $\beta$ increases.\label{fig7}}}
\end{figure}

One can show that the Ricci and Kretschmann scalar have the asymptotic values
\ben
\lim_{y\to\pm\infty}R\!\!&=&\!\!\frac{80}{9}(1+3\beta)^2[1-(\alpha\kappa)^2]\,,\nn
\lim_{y\to\pm\infty}K\!\!&=&\!\!\frac{640}{81}(1+3\beta)^4[1-(\alpha\kappa)^2]^2\,.\nonumber
\een
On the other hand, at the origin we get that
\ben
\lim_{y\to0}R\!\!&=&\!\!-\frac{16}{3}(1+3\beta)(1+\alpha\kappa)\,,\nn
\lim_{y\to0}K\!\!&=&\!\!\frac{64}{9}(1+3\beta)^2(1+\alpha\kappa)^2\,.\nonumber
\een
In addition, the trace $T$ behaves as
\ben
\lim_{y\to\pm\infty}T\!\!&=&\!\!-\frac{20}{3\lambda^2}(1+5\beta)[1-(\alpha\kappa)^2]\,,\nn
\lim_{y\to0}T\!\!&=&\!\!\frac{(1+\alpha\kappa)[4+9\beta(3+5\beta)-\alpha\beta]}{(1+5\beta)(1+3\beta)}\,.\nonumber
\een
They develop no divergence for the allowed values of parameters. The energy density is represented in Fig. \ref{fig7} for the same values of $\alpha$ and $\beta$ used to calculate the warp factor.

\begin{figure}[!htb]
    \begin{center}
        \includegraphics[scale=0.6]{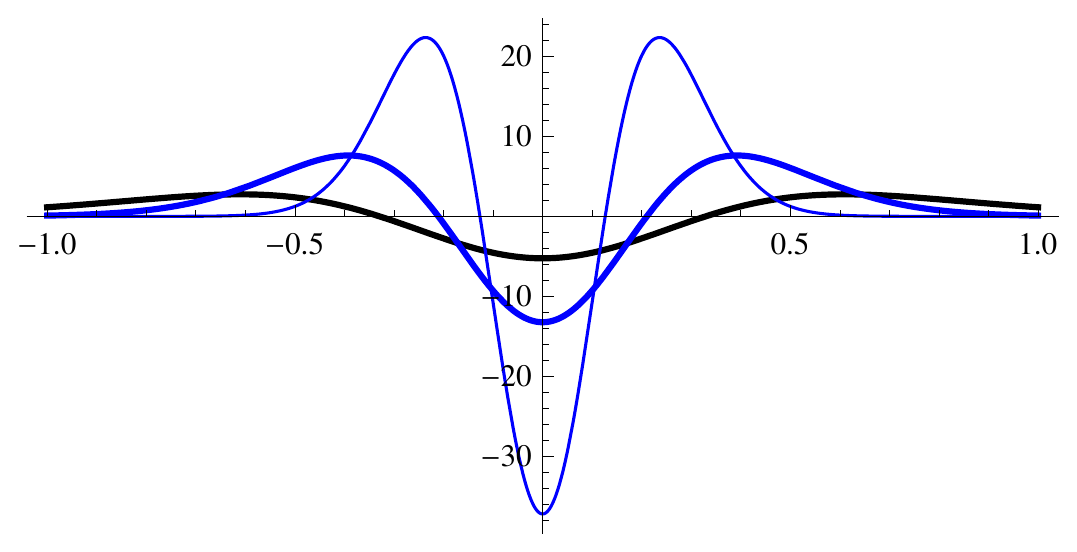}
    \end{center}
    \vspace{-0.7cm}
    \caption{\small{Potential of gravitational mode for $\alpha\!=\!1/4$ and $\beta\!=\!1$ (black line), $\beta\!=\!3$ and $9$ (blue line). The line thickness decrease as $\beta$ increases.\label{fig8}}}
\end{figure}

\begin{figure}[!htb]
    \begin{center}
        \includegraphics[scale=0.6]{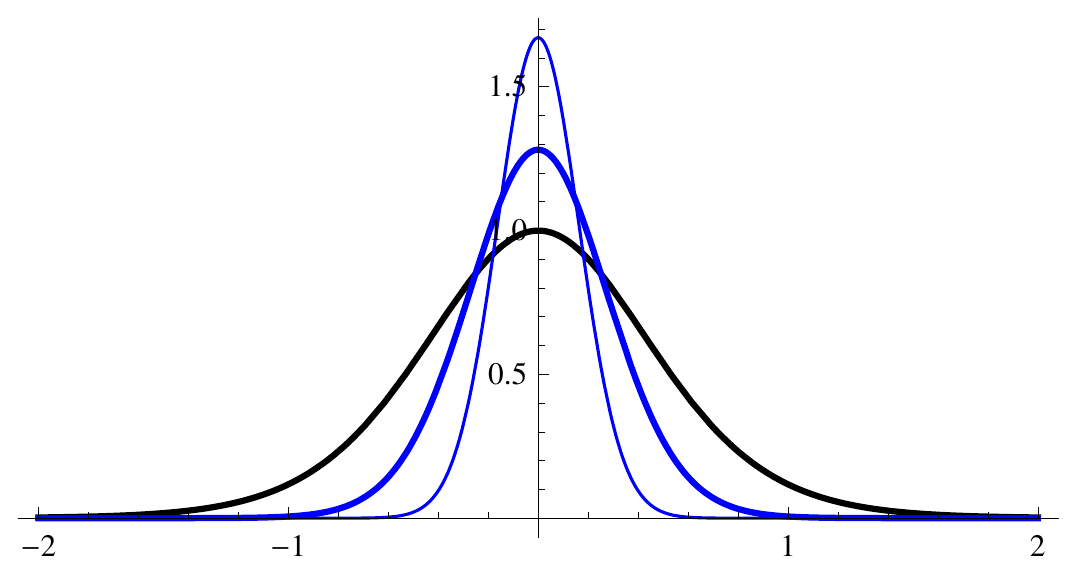}
    \end{center}
    \vspace{-0.7cm}
    \caption{\small{Zero mode for $\alpha\!=\!1/4$ and $\beta\!=\!1$ (black line), $\beta\!=\!3$ and $9$ (blue line). The line thickness decrease as $\beta$ increases.\label{fig9}}}
\end{figure}

In Figs. \ref{fig8} and \ref{fig9} we display the gravitational potential and zero mode obtained for the same values of $\alpha$ and $\beta$ described above, respectively. In all figures the line thickness decrease as $\beta$ increases.
\\

{\bf 3.2. Second Model.} --
Let us now consider another modification of the theory, assuming that the functions $G(T)=\beta T$ and $F(R)$ has now the form $ 
F(R)=R-f(\beta)\,R^2$,
where $f(\beta)=(1\!+\!3\beta)/308$ is used to simplify the equations of the model. In this case, the Einstein equations \eqref{eq8} become
\bes\label{eq28}
\ben
\!\!\!&&\!\!\!A^{\prime\prime}-\frac{50}{231}(1+3\beta)A^{\prime 2}A^{\prime\prime}-\frac{32}{231}(1+3\beta)A^{\prime\prime 2}+\nn
\!\!\!&&\!\!\!-\frac{8}{77}(1+3\beta)A^{\prime}A^{\prime\prime\prime}-\frac{4}{231}(1+3\beta)A^{\prime\prime\prime\prime}\nn
\!\!\!&&\!\!\!=-\frac23(1+3\beta)\phi^{\prime 2}+\frac23\alpha(1+4\beta) \phi^{\prime}\,,\label{eq28a}\\
\!\!\!&&\!\!\!A^{\prime2}\!-\!\frac{5}{231}(1+3\beta)A^{\prime 4}\!-\!\frac{32}{231}(1+3\beta)A^{\prime 2}A^{\prime \prime}+\nn
\!\!\!&&\!\!\!+\frac{4}{231}(1+3\beta)A^{\prime\prime 2}\!-\!\frac{8}{231}(1+3\beta)A^{\prime}A^{\prime\prime\prime}\nn
\!\!\!&&\!\!\!=\frac16(1\!+\!3\beta)\phi^{\prime 2}-\frac13(1\!+\!5\beta)V(\phi)\,.\label{eq28b}
\een
\ees
Since we are interested in thick brane solutions, as suggested in \cite{Bazeia:2013uva}, we chose to set an explicit form for the warp function $A(y)$ as $A(y)=\ln \left(\sech(y)\right)\,.$
This is standard bell-shaped solution for the warp factor, and it allows that we rewrite the relation \eqref{eq28a} in the form
\ben\label{eq30}
\sech^4(y)-\frac{9(38\beta\!-\!13)}{154(1+3\beta)}\sech^2(y)=\phi^{\prime 2}-\alpha\kappa \phi^{\prime}\,,
\een
where in this new model, $\kappa=(1+4\beta)/(1+3\beta)$, as used before. This equation has analytical solution for an adequate choice of parameter $\alpha$. Specifically, if $\alpha=9(38\beta-13)/(616\beta-154)$ we get
$\phi(y)=\tanh(y)\,.$ 
With this, the potential can be obtained by the Eq. \eqref{eq28b} in the form
\begin{equation}\label{eq32}
    V(\phi)=\frac{1}{154\lambda}\Big[69-\frac{6(116+117\beta)}{1+3\beta}\phi^2+175\phi^4\Big]\,,
\end{equation}
where $\lambda=(1+5\beta)/(1+3\beta)$, as used before. The five-dimensional cosmological constant is $\Lambda_5=(15\beta-226)/(385\beta+77)$ showing that the bulk is asymptotically Minkowski for $\beta=226/15$ and $AdS_5$ for $-1/5<\!\beta\!<\!226/15$.

The energy density can be written as
\ben\label{eq33}
\rho(y)=\frac{r_1S^2+r_2S^4+r_3S^6}{77(1+5\beta)}\,,
\een
where $S=\sech(y)$ and $r_1=15\beta-226$, $r_2=(229-1975\beta+3102\beta^2)/(2-8\beta)$ and $r_3=7(18+65\beta)$. The energy of the brane is then obtained as a function of the $\beta$ parameter in the form
\ben\label{eq34}
E=\frac{2(50\beta^2+3518\beta-1237)}{1155(1+5\beta)(1-4\beta)}\,.
\een
Note that the energy vanishes for $\beta=\bar\beta=(3\sqrt{350659}-1759)/50=0.350$, and it is positive for $1/4<\beta<\bar{\beta}$.

\begin{figure}[!htb]
    \begin{center}
        \includegraphics[scale=0.6]{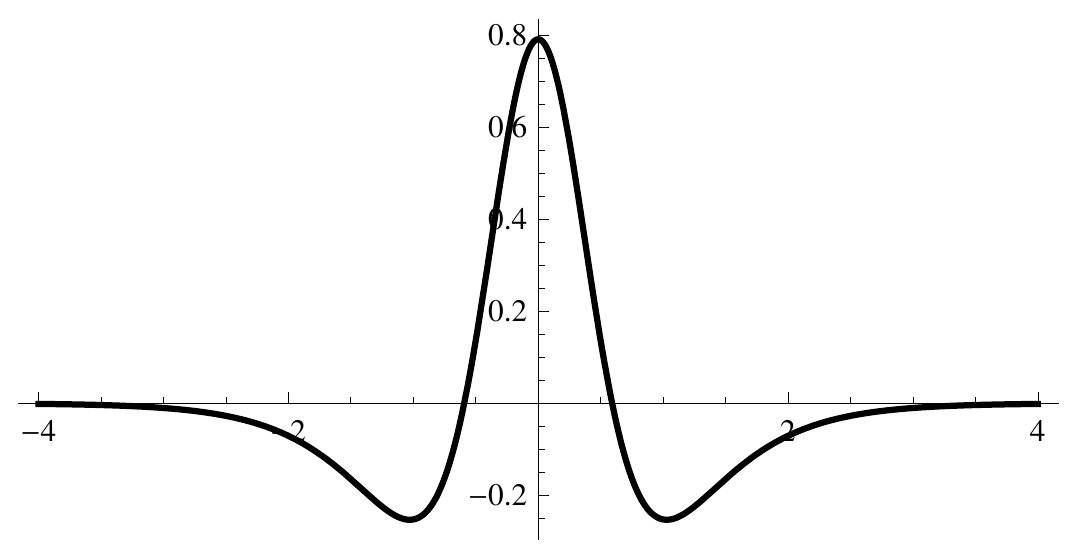}
    \end{center}
    \vspace{-0.7cm}
    \caption{\small{Energy density for $\beta=0.350$.}
    \label{fig10}}
\end{figure}

\begin{figure}[!htb]
    \begin{center}
        \includegraphics[scale=0.6]{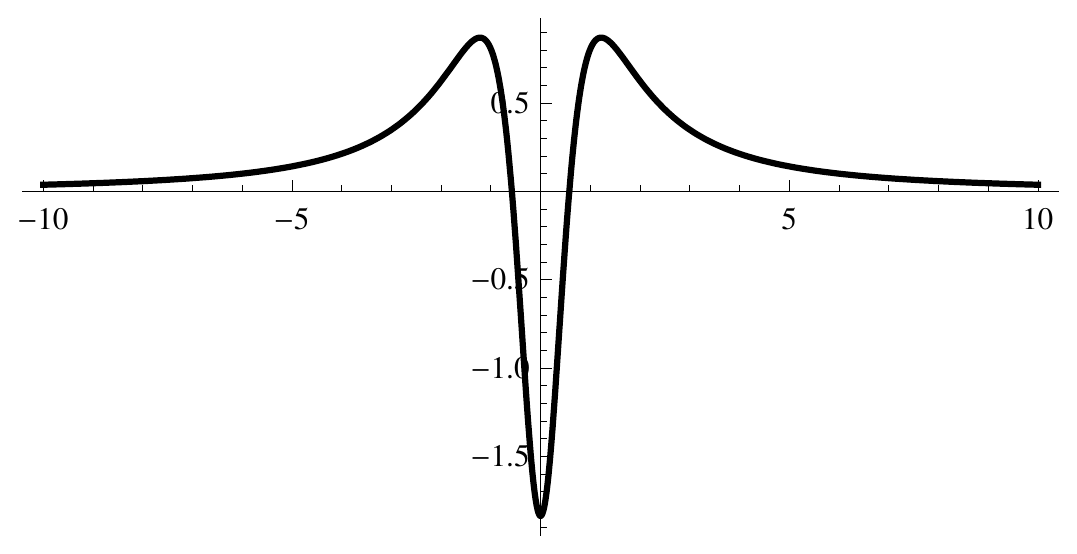}
    \end{center}
    \vspace{-0.7cm}
    \caption{\small{Potential of gravitational mode for $\beta\!=\!0.350$.\label{fig11}}}
\end{figure}

\begin{figure}[!htb]
    \begin{center}
        \includegraphics[scale=0.58]{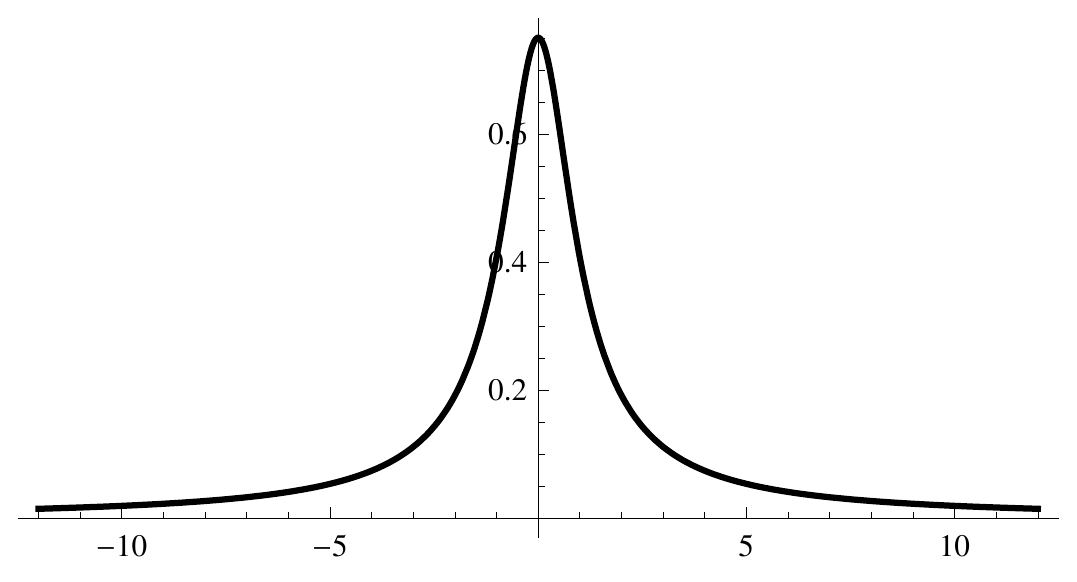}
    \end{center}
    \vspace{-0.7cm}
    \caption{\small{Zero mode for $\beta\!=\!0.350$.\label{fig12}}}
\end{figure}

In this model we can find the potential of gravitational mode and the zero mode in an explicit form. However, we do not present the results explicitly since the equations are long and awkward and they express no new possibilities; however, we must impose that $\beta<67/30$ in order to keep the results real. Next, we display the energy density, zero mode and potential of gravitational mode in the case of brane with zero energy in Figs. \ref{fig10}, \ref{fig11} and \ref{fig12}; the case of positive energy gives similar results. 


Here both the Ricci and Kretschmann scalars are non-divergent by construction, since they only depend on the warp function. However, we have to analyze the behavior of the trace $T$. Similarly to the models studied in the previous section we can show that
\ben
\lim_{y\to\pm\infty}T\!\!&=&\!\!\frac{(15\beta-226)}{77(1+5\beta)}\,,\nn
\lim_{y\to0}T\!\!&=&\!\!-\frac{3(28+135\beta-232\beta^2)}{77(1+5\beta)(1-4\beta)}\,.\nonumber
\een
One can verify that this result does not diverge in the region of values $1/4<\beta\leq\bar\beta$, which includes the bulk asymptotically $AdS_5$ and Minkowsky.
\\

{\bf  4. Conclusion.}
In this work we obtained brane solutions in generalized braneworld models. We studied modifications both in the gravitational and in the dynamics of the scalar field, through the inclusion of the cuscuton term, the trace of the stress-energy tensor and a generalized function of the Ricci scalar. Several characteristics of the models were analyzed with care, including stability, warp factor, energy density and the zero mode, in terms of the specific parameters that define each one of the several models included in the work.

In the first situation studied, we found two distinct stable solutions for models with cuscuton in the presence of a trace term in the form $G(T)=\beta T$, using the standard relation for the Ricci scalar, $F(R)=R$. In this case, we verified that the parameter $\beta$ contributes to shrink the brane towards its center. We also noticed that the proposed generalizations do not destabilize the geometric degrees of freedom of the brane, being the model stable in this perspective. In the second situation, we included a quadratic term for the Ricci scalar in the form $F(R)=R-f(\beta)R^2$, also in the presence of a linear trace term. In this model we described analytical solutions for an adequate choice of parameter $\alpha$ and we controlled the stability by adjusting the $\beta$ parameter. In each model, we verified that there are no divergences in the Ricci and Kretschmann scalars and in trace of the stress-energy tensor for the allowed range of parameters.

An interesting result is that the presence of the cuscuton term do not contribute to split the warp factor in braneworld models of a single-field generalized via the inclusion of a linear term in the trace of stress-energy tensor and/or quadratic term of the Ricci scalar. We verified that the proposed generalizations cause no destabilization of the geometric degrees of freedom and also, unlike the investigation considered in \cite {Bazeia:2021jok} with the cuscuton dynamics, no split was found in the warp factor. Further study is still required to better understand how the trace $T$ modifies the warp factor of the brane.  Moreover, it turns out that the additional parameters presented in this paper can be used in other contexts where the cuscuton seems to play important role, for instance, in the case \cite{Cuscuton}, and also in \cite{C} in which the authors deal with cosmology in cuscuton gravity, searching for an exact solution describing accelerated four-dimensional universe with a stable extra dimension. In this context, we can add another scalar field, as done in \cite{Bazeia:2021jok}, to investigate how the extra field may change the braneworld configurations found in the present work. This and other related issues are now under investigation, to be reported elsewhere.

Work supported by Paraiba State Research Foundation, Grant 0015/2019. DB also thanks CNPq, Grants 303469/2019-6 and 404913/2018-0, for partial financial support.

    
\end{document}